# Charge-independent mass spectrometry of single virus capsids above

# 100MDa with nanomechanical resonators

**Authors:**

Sergio Dominguez-Medina[1,2,3], Shawn Fostner[4], Martial Defoort[4], Marc Sansa[4], Ann-Kathrin Stark[1,2,3], Mohammad Abdul Halim[1,2,3], Emeline Vernhes[5,•], Marc Gely[4], Guillaume Jourdan[4], Thomas Alava[4], Pascale Boulanger[5], Christophe Masselon[1,2,3,†], Sébastien Hentz[4,‡]

**Affiliations:**

1. Université Grenoble-Alpes, F-38000 Grenoble, France
2. CEA, BIG, Biologie à Grande Echelle, F-38054 Grenoble, France
3. Inserm, Unité 1038, F-38054 Grenoble, France
4. Univ. Grenoble Alpes, CEA, LETI, 38000 Grenoble, France
5. Institute for Integrative Biology of the Cell (I2BC), CEA, CNRS, Univ Paris-Sud, Université Paris-Saclay, 91198, Gif sur Yvette cedex, France
• Present address : Laboratoire de Microbiologie et Génétique Moléculaires (LMGM), Centre de Biologie Intégrative (CBI), CNRS, Université Paul Sabatier, Université de Toulouse, 118 Route de Narbonne, 31062 Toulouse Cedex, France

† christophe.masselon@cea.fr
‡ sebastien.hentz@cea.fr

**Abstract:**

*Most technologies, including conventional mass spectrometry, struggle to measure the mass of particles in the MDa to GDa range. Although this mass range appears optimal for nanomechanical resonators, early nanomechanical-MS systems suffered from prohibitive sample loss, extended analysis time or inadequate resolution. Here, we report on a novel system architecture combining nebulization of the analytes from solution, their efficient transfer and focusing without relying on electromagnetic fields, and the mass measurements of individual particles using nanomechanical resonator arrays. This system determined the mass distribution of ~30 MDa polystyrene nanoparticles with a detection efficiency 6 orders of magnitude higher than previous*

*nanomechanical-MS systems with ion guides, and successfully performed the highest molecular mass measurement to date with less than 1 picomole of bacteriophage T5 105 MDa viral capsids.*

**One sentence summary:** *A new mass spectrometer architecture using nanomechanical resonators allows efficient analysis of individual nanoparticles and 105MDa virus capsids.*

Mass was one of the earliest physical properties to be measured. Today, objects can be weighed *individually* over a range from the pg (with piezoelectric resonators (*1*)) to the ton. Thirty orders of magnitude lower on the mass scale, mass spectrometry (MS) uses ionization, electromagnetic fields to manipulate ions, and *ensemble averaging* of mass-to-charge ratios to identify species based on their specific molecular mass. Early spectrometers operated in the Da to 1-kDa range (1 Da=1.66 $10^{-27}$ kg = 1.66 yg), and since the advent of soft ionization techniques (*2, 3*), commercial spectrometers have performed routine proteomics analysis of heavier species such as proteins and peptides in the kDa to 100-kDa range. More recently, MS research has devoted its efforts to expanding the mass range accessible with conventional techniques, up to the 10-MDa range (*4*). Alternative MS techniques, such as Charge Detection Mass Spectrometry (CDMS) (*5, 6*) can analyze single ions with masses up to a GDa. Nevertheless, to date, no commercially-available MS instruments can measure masses above a MDa. Nanomechanical resonators measure the mass of individual particles accreting on their surface (*7*). Because the frequency-to-mass relationship scales with the resonator's characteristic dimension $d^{-4}$, nanomechanical system researchers have engaged in a race to measure the smallest detectable mass with ever smaller resonators, eventually using processes that are far removed from microelectronics standards (*8– 12*). It is interesting to note that this was driven by the hope of eventually competing with commercial MS down to a Da; meanwhile, the MS community has worked in the opposite direction. Moreover, nanomechanical resonators fabricated by scalable silicon processes have been ideally suited for analysis of masses in the MDa to GDa range (*7*). Unfortunately, early attempts to perform nanomechanical-MS were hampered by a combination of losses associated with ionization yield and ion transfer, and the small cross-section (typically a few $\mu m^2$) of the capture area on nanoresonators. These issues led to prohibitively long analysis times and excessive sample consumption (*13*). An important milestone on the path toward alternative system architectures was recently reached, with the demonstration of nanomechanical-MS of metallic nanoparticles of any charge state (ionized or neutral) (*7*).

To achieve the full potential of nanomechanical-MS in the MDa to GDa range, we report here on a novel system architecture specifically designed for this mass range (see Fig. 1). This architecture features high efficiency and excellent resolution by circumventing the requirement for ionization; it relies on nebulization of analytes from solution at atmospheric pressure, exploits the particles' inertia to efficiently transfer and focus them without any need for electromagnetic fields, and determines the mass of individual particles using arrays of nanomechanical resonators.

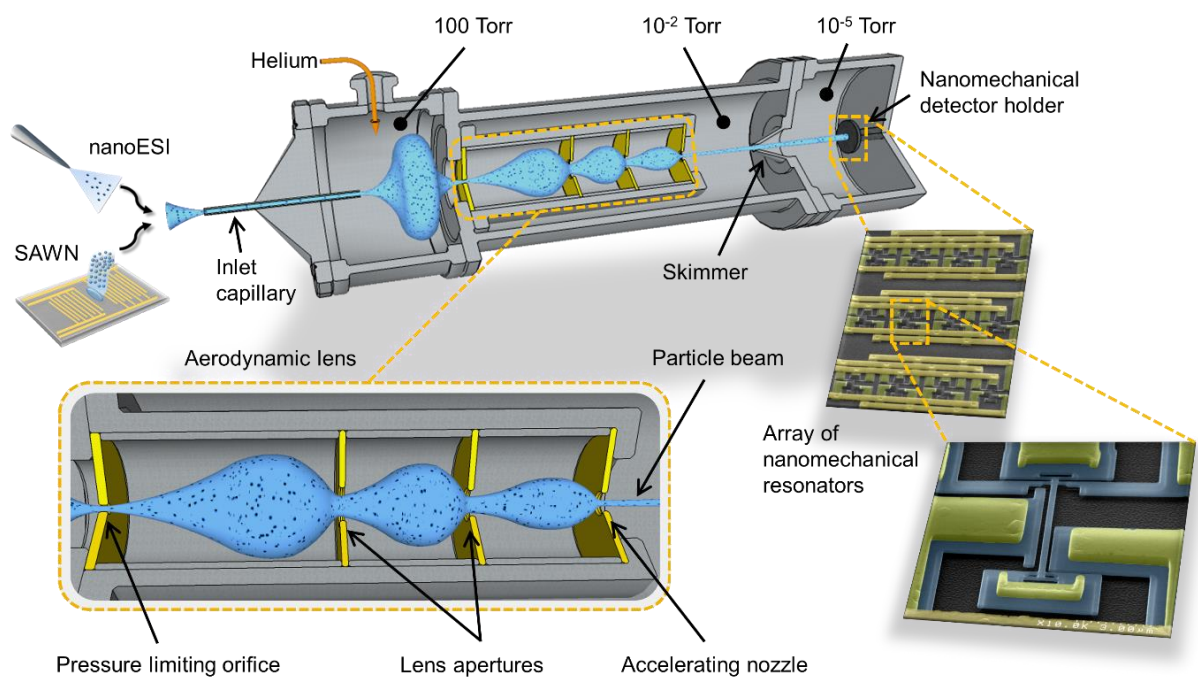

**Fig. 1. High-transmission system architecture for nanomechanical resonator-based charge-independent single particle mass sensing.** The setup consists of three chambers with decreasing pressures. Analytes in solution are nebulized by SAWN or nano-ESI, and aspirated through a heated metal capillary inlet at atmospheric pressure. An aerodynamic lens focuses the particle stream (shaded blue area) which is then transferred onto an array of frequency-addressed nanomechanical resonators. Left inset: simplified view of the aerodynamic lens. Right inset: SEM image of a portion of an array showing 12 out of 20 resonators and magnified SEM image of a single resonator with its metallic layer and silicon material, falsely colored in yellow and gray, respectively. Resonator dimensions: 160 nm (thickness), 300 nm (width) and 7 to 10 µm (length).

Nebulization of particles was performed by Surface Acoustic Wave Nebulization (SAWN), or by Electro Spray Ionization (ESI). With SAWN, a liquid sample deposited on a piezoelectric surface is nebulized as a thin mist upon propagation of an acoustic wave across the surface (*14*). Compared to ESI, which produces a rapidly expanding jet due to Coulombic repulsion, SAWN produces droplets with lower kinetic energy, resulting in very efficient uptake into the system's inlet capillary. The low ionization efficiency of SAWN compared to ESI (*14*) is not problematic here as our system can analyze both ions and neutrals. On the other hand, ESI conditions have been extensively optimized for numerous species of interest, including viruses (*4*).

Due to the small capture area of nanoresonators, one of the key elements for efficient particle detection is the particle-beam-to-detector size ratio: we used an aerodynamic lens (or aerolens) to focus the particles into a narrow beam by exploiting their inertia against the natural tendency for the gas flow to expand. An aerolens is composed of a pressure-limiting orifice followed by a series of apertures between relaxation volumes (Fig. 1). Ours was designed and optimized for 10-100 nm-diameter particles with a density close to 1 g.cm$^{-3}$, following previously published guidelines (*15*). Compared to conventional ion guides, which struggle against inertia to maintain a stable trajectory for heavy particles, the performance of the aerolens improves with increasing mass (*16*).

The detector consisted of 20 nanomechanical resonators arranged in a 4-by-5 array covering a 50 x 230 μm$^2$ area. Individual resonators were simply-supported beams with electrostatic actuation and differential piezoresistive readout fabricated using large-scale silicon processes (*17*). Particles landing on the vibrating part of a resonator add to its total mass $M$ and cause its resonance frequency $f$ to down-shift ($f \propto \sqrt[-1]{M}$). As these frequency shifts also depend on the landing position on the resonator's surface, the frequencies of two resonance modes were monitored simultaneously to resolve the two unknowns (*i.e.* mass and position) (*7*, *13*, *18*). The effect of particle stiffness on frequency was negligible in our case (*19*). With an array of 20 resonators multiplexed in time, an increase of more than an order of magnitude in capture cross-section (and hence detection efficiency) could be obtained without degrading the mass resolution (*20*, *21*), in our case a few 100 kDa (*22*).

We first demonstrated the capacities of this novel architecture using ~45nm diameter polystyrene nanoparticles (NIST). The transmission and focusing performance of our system was characterized by exposing silicon targets to the particle beam produced by SAWN (Fig. 2a). The transmission efficiency defined as the ratio of the number of particles on the targets (determined by SEM observation) to the number of particles in the nebulized solution ranged from 3.5 to 7.6%. This transmission was 4 to 5 orders of magnitude greater than that achieved with previously reported architectures combining ion guides with nanomechanical resonators (*13*). The measured full width at half maximum (FWHM) of the beam profile at the detector position was significantly smaller than both the aerolens outlet (1.9mm) and the skimmer's orifice (2mm) (Fig. 2b). We deduced a solid angle of less than 1°, indicating near-perfect collimation of the particle beam delivered by our aerodynamic lens.

After aligning the nanomechanical array with the ~1.5mm diameter particle beam, NIST polystyrene nanoparticles from a $1.7 \ 10^7$-NP·µl$^{-1}$ (28.2 pM) solution were measured with our system. Fig. 2c shows the frequency time-traces obtained for one resonance mode (similar traces were obtained for the second mode) of all resonators in the array over 10 minutes of acquisition, showing downward frequency jumps corresponding to individual particle landing "events". An event rate ranging from 0.3 to 1.8 NP·min$^{-1}$·resonator$^{-1}$ was recorded over 128 minutes of nebulization at an average flow rate of 2 µl·min$^{-1}$. The event rate was influenced by detector alignment and nebulization conditions. The resonance frequency for a given resonator in the array was sampled approximately every 200 ms, *i.e.,* more than 2 orders of magnitude faster than the highest event rate. As a result, the probability of two particles landing on the same resonator within this sampling time was negligible. The particles average size and standard size deviation $\sigma$ determined beforehand by SEM (45 ±3nm and $\sigma = 13nm$) were in line with the NIST specifications (46 ±2nm and $\sigma = 7nm$). Translating these size measurements to mass, we could expect a very broad mass distribution, with a central mass ranging from 28 to 36 MDa and a standard mass deviation of about 15 MDa (*22*). Fig. 2d shows the mass histogram constructed from the nanomechanical measurements of 173 individual particles. The normal fit of this histogram (29.5 MDa) was in good agreement with the expected mass range. The absolute mass error in our measurement was mainly due to errors in determination of the resonators' effective mass. We estimated this absolute error to be between 1 and 2% (300 to 600 kDa) (*22*), which was well below the uncertainty in central mass expected from size measurements (4 to 6 MDa). Moreover, the standard deviation of the mass distribution expected from size measurement ranged from 15 (NIST) to 25 MDa (SEM), whereas we measured a 17-MDa deviation. This difference was not due to our measurement noise level: the mass resolution for each measured particle could be inferred from the frequency noise and landing position on the resonator (*22*), and this estimation yielded an average mass resolution of 600 kDa for the 173 events. Our mass measurement was thus consistent with expectations from size measurements. The total amount of nanoparticles consumed during this measurement was only 7.3 femtomoles (4.4 $10^9$ particles), and the detection efficiency was one particle per 2.2 $10^7$ particles in solution. These numbers are 6 orders of magnitude better than previous nanomechanical resonator-based systems using ion guides (*13*). They also compare well with typical figures for routine measurements performed in a vastly different mass range with commercial mass spectrometer, making our nanomechanical-MS system compatible with "real-life" experiments.

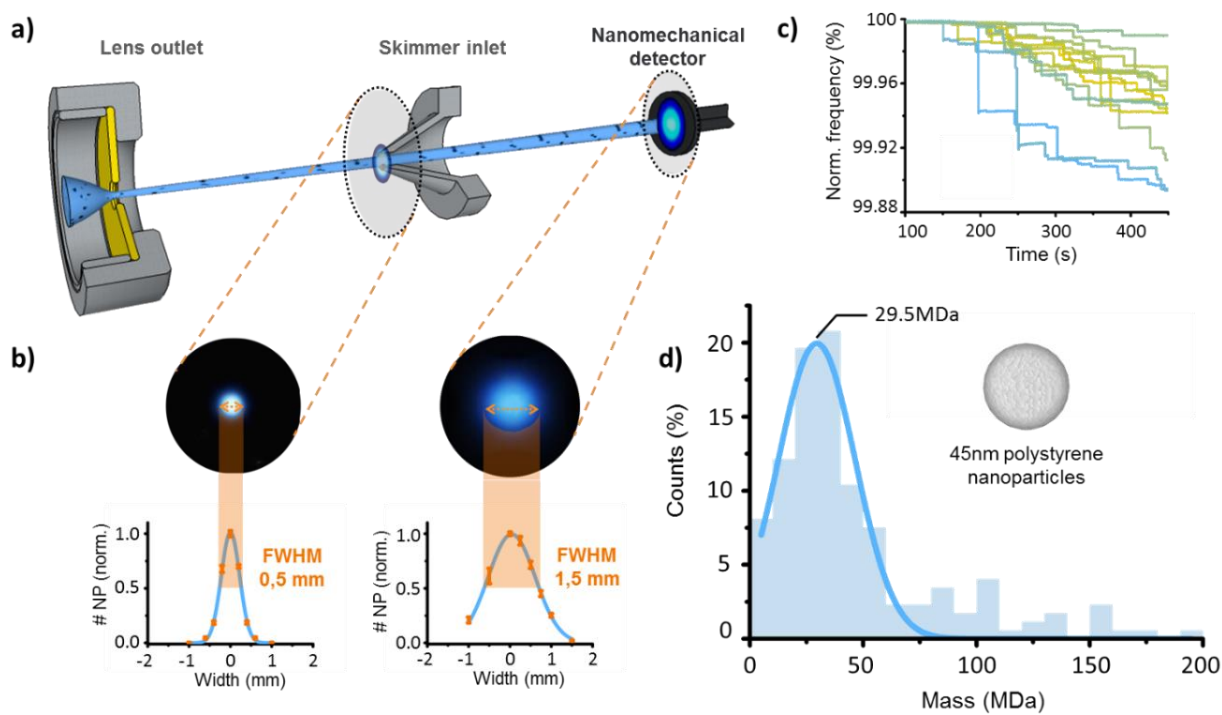

**Fig. 2. Characterization of SAWN-based transmission and focusing efficiency, and high throughput mass spectrum of polystyrene nanoparticles.** a) Diagram showing how the particle beam was characterized by placing silicon targets at the skimmer (~4cm from aerolens outlet) and nanomechanical detector (~8cm) positions. b) Optical images of the targets observed after deposition of NIST polystyrene particles, with respective Gaussian fit to horizontal sections of the beam profile in particles per μm² (normalized scale) as measured on a series of SEM images. c) Normalized raw frequency traces as a function of time for the fundamental mode of 15 nanoresonators exposed to the particle beam produced by SAWN. Each color represents a different nanomechanical device, and each step corresponds to a particle landing event. d) Accumulated histogram of mass measurements of the same polystyrene nanoparticles for a nanoresonator array exposed to the particle beam, fitted to a normal distribution.

Going further, we performed measurements of a biological particle: the capsid of bacteriophage T5. Bacteriophage T5 is a *siphoviridae* family member which infects *Escherichia coli* bacteria. It is composed of a ~90-nm icosahedral capsid connected to a 250-nm tail which plays a role in host cell recognition and genome delivery. The capsid is composed of 775 copies of the major capsid protein pb8 and 12 copies of the portal protein pb7. Well-controlled capsid assembly and expansion without ("empty" capsids (*23*)) and with ("filled" capsids (*24*)) genome packaging inside the capsid have been demonstrated. Once loaded with its 121.75 kbp dsDNA viral genome, the capsid is completed by the head completion protein p144, which constitutes the docking site for tail attachment. Phage T5 is one of the only viruses with such large genome content amenable to in vitro studies. Unlike polystyrene nanoparticles, empty and filled capsids have distinct molecular masses, calculated to be 26.0 MDa and 105.4 MDa, respectively (see Fig. 3).

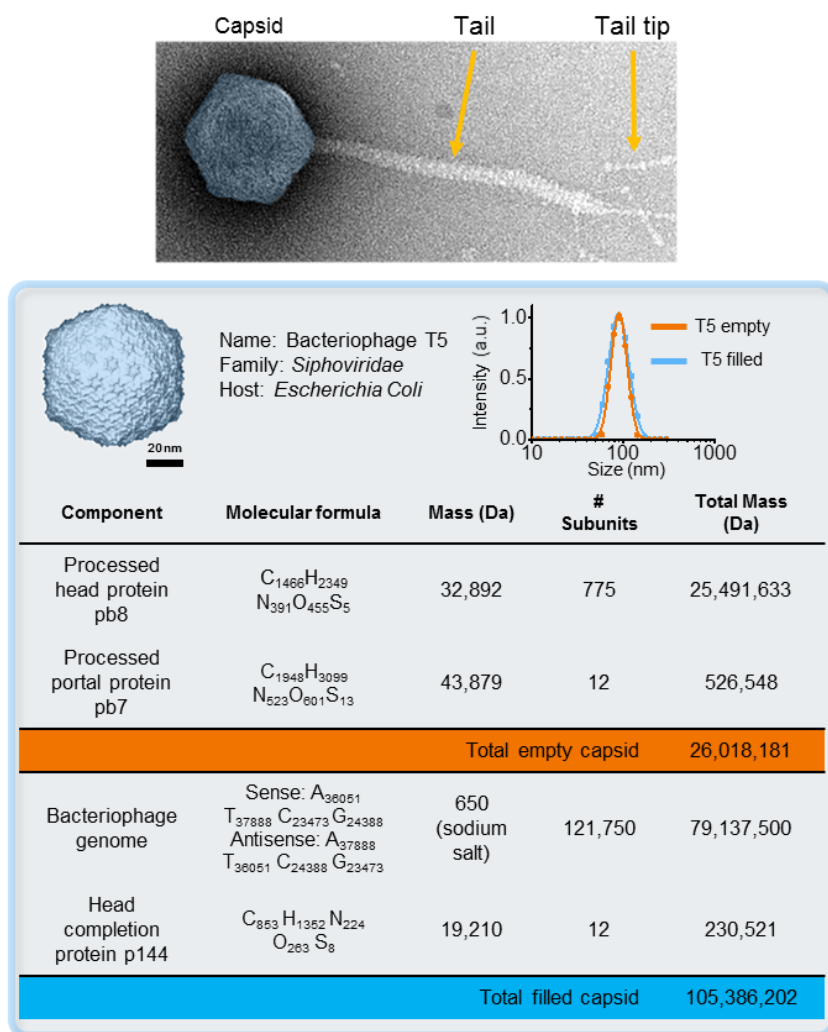

| Component | Molecular formula | Mass (Da) | # Subunits | Total Mass (Da) |
|---|---|---|---|---|
| Processed head protein pb8 | $C_{1466}H_{2349}N_{391}O_{455}S_5$ | 32,892 | 775 | 25,491,633 |
| Processed portal protein pb7 | $C_{1948}H_{3099}N_{523}O_{601}S_{13}$ | 43,879 | 12 | 526,548 |
| | | | Total empty capsid | 26,018,181 |
| Bacteriophage genome | Sense: $A_{36051}T_{37888}C_{23473}G_{24388}$ Antisense: $A_{37888}T_{36051}C_{24388}G_{23473}$ | 650 (sodium salt) | 121,750 | 79,137,500 |
| Head completion protein p144 | $C_{853}H_{1362}N_{224}O_{263}S_8$ | 19,210 | 12 | 230,521 |
| | | | Total filled capsid | 105,386,202 |

Name: Bacteriophage T5
Family: *Siphoviridae*
Host: *Escherichia Coli*

20nm

**Fig. 3. Structure and size of the bacteriophage T5 capsid.** Top: negatively-stained electron microscopy image of the native bacteriophage T5. The capsid is falsely colored in blue. Bottom: 3D reconstruction of the assembled capsid measured here (adapted from (*23*)). Scale bar 20 nm. Dynamic light scattering size measurements of empty and filled capsids: both capsids display very similar sizes while their mass differs widely. The table shows the components of the capsid, with theoretical molecular mass calculations for both types of capsid.

To nebulize the capsids, we chose to use nano-ESI as it had an excellent track record for native MS of high mass biological analytes (*4*). Fig. 4a shows the mass distribution obtained for empty capsids. The spectrum displayed a bimodal distribution, which was fitted by two Gaussian peaks. The main peak had a central mass of 25.9 MDa ($\sigma = 850 kDa$), which was in agreement with the calculated 26.02 MDa; the second peak had a mass of 28.4 MDa ($\sigma = 1.3 MDa$). From the mass resolutions obtained for each single particle in the main peak, it was possible to deduce a 1.6 MDa FWHM due to the instrument alone (*i.e.,* considering negligible sample heterogeneity) (*22*). This corresponds to a resolution of 16 ($M/\Delta M_{FWHM}$), which is comparable to the rare measurements in this mass range (*6*). Moreover, it suggests that the two peaks could be ascribed to two distinct capsid populations truly present in the sample. This was further substantiated by a differential scanning calorimetry experiment (*22*) showing two denaturation profiles. This could be attributed to the presence of a remaining short DNA fragment associated to some T5 capsids during purification. The total amount of sample consumed for this analysis was 580 femtomoles (1 particle detected per 3.57 $10^9$ in solution). This result confirms the advantages of using our architecture even with the lower uptake efficiency of nano-ESI compared to SAWN.

We then nebulized filled capsids and measured their mass distribution (Fig. 4b). A clear peak emerged at a mass of 106.2 MDa ($\sigma = 3.2 MDa$), as well as two secondary peaks with central masses of 113.7MDa ($\sigma = 1.7 MDa$) and 118.7MDa ($\sigma = 4 MDa$). The central mass of the main peak was in excellent agreement (within ~1%) with the filled capsid calculated mass (105.4 MDa). The slight remaining discrepancy could be due to the electrospray process which reportedly yields masses slightly higher than theoretical estimates (*25*). Individual polyhedral particles of the expected size (~93 nm) were clearly discernible on the nanomechanical resonators observed under SEM (*22*). Remarkably, most capsids not only maintained their structure during nebulization and transfer across the vacuum interface and the aerolens, but they also survived landing on the nanoresonator surface at nearly super-sonic speeds, further suggesting the main peak could be associated to filled capsids. The secondary peaks at 113.7 MDa and 118.7MDa could be due to the presence of fragments of damaged capsids adhering to intact capsids. The total amount of sample consumed for this analysis was 215 femtomole (1 particle detected per 6 $10^8$ in solution). This six-fold improvement in detection efficiency compared to empty capsids is likely due to improved focusing due to the higher masses of the particles. Remarkably, Fig. 4c and 4d show that the average mass resolution per detected filled capsid was the same as that obtained with empty capsids (~640kDa). This is the first experimental evidence of this much-vaunted feature of nanomechanical mass sensing over such a large mass range, confirming that nanomechanical MS has a higher resolution ($M/\Delta M_{FWHM}$) at higher masses. From the mass resolutions obtained for each single capsid in the main peak, we determined a 1.4 MDa FWHM due to instrument alone (*i.e.,* in the absence of sample heterogeneity), which corresponds to a resolution of 75 at 106 MDa. This measurement of the mass of bacteriophage T5 capsids is, to the best of our knowledge, the largest defined molecular mass ever measured, and the resolution is the highest reported for single particle measurements above 10 MDa (*6*).

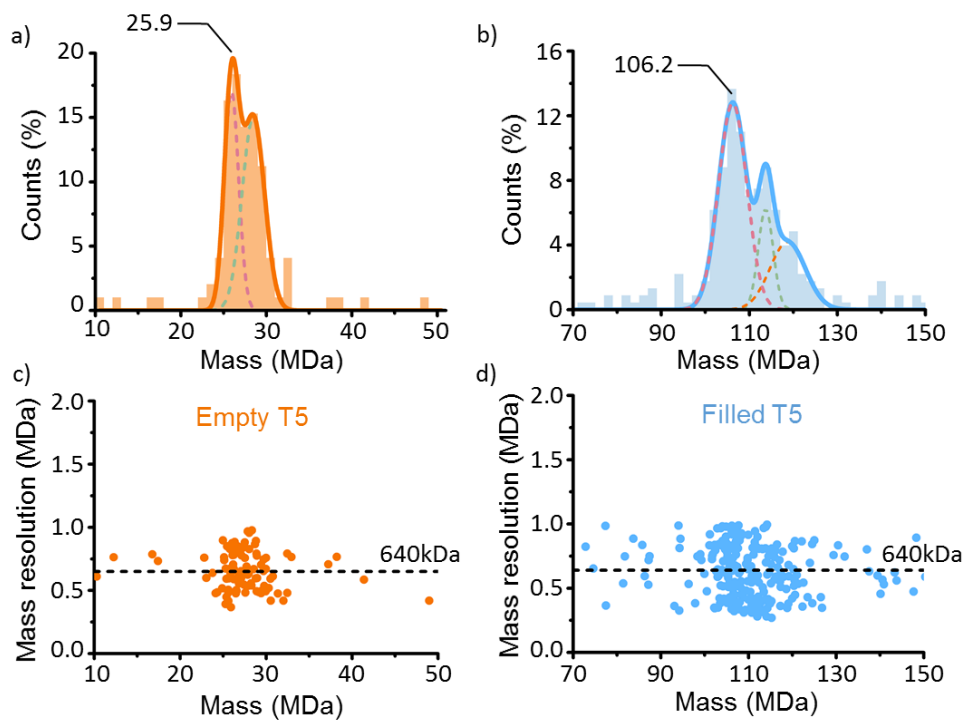

**Fig. 4. Molecular mass measurement of empty and filled bacteriophage T5 capsids.** (a-b) Accumulated mass histograms of empty (a) and filled (b) capsids of bacteriophage T5 nebulized using nano-ESI (with 1MDa and 2MDa bin size, respectively), fitted to 2 and 3 normal distributions, respectively. Mass resolution per particle for empty (c) and filled capsids (d).

The mass spectrometer architecture presented here is one of the very rare technologies that can operate in the MDa to GDa range and beyond, and has the unique capability to analyze ionized or neutral species. It overcomes many of the limitations associated with earlier nanomechanical resonator-based systems, the main one of which was inefficient detection. In our system, simple nebulization techniques can be used, leading to efficient uptake as ionization yield or Taylor cone expansion do not come into play. Moreover, these techniques may be less prone to dissociating non-covalently bound supramolecular analytes. Rather than being counteracted by electromagnetic fields, the inertia of massive particles is exploited for efficient guiding and focusing using an aerodynamic lens. Finally, nanomechanical resonators directly measure the inertial mass of individual analytes, avoiding separate measurements of *m/z* and charge. In this report, experiments were performed with concentrations and sample quantities that were close to those typically used in MS experiments, and the total duration of experiments was only a few hours. As a result, routine analysis in an extremely difficult-to-access mass range is now at hand. Using SAWN, a few % of particles in solution reach the detector chamber, a level which exceeds that obtained in conventional MS by orders of magnitude. However, most particles remained undetected as the detector-to-beam area ratio was only around $10^{-6}$. In the very near future, thanks to the very large-scale integration of thousands of nanoresonators with CMOS (Complementary Metal Oxide Semiconductor) and working in parallel (*26*), this type of measurement will take only minutes with extremely low amounts of sample.

We have demonstrated that our system operates in a range where objects with wide molecular variability and objects with well-defined molecular mass overlap. Nanomedicine for therapeutic or imaging approaches can harness the power of nanoparticles to control the biodistribution or enhance the efficacy of a drug or biologic, and interest in this field is growing rapidly (*27*). However, these techniques are not yet commonplace, as only few instances have received FDA-approval, due in part to the lack of available characterization methods (*27*). The system we propose has the features required to measure individual particle mass and determine drug loading heterogeneity. We analyzed the mass of phage virus capsids above 100 MDa with adequate resolution to provide valuable information on their protein- and DNA content as well as their assembly process. Applications of phage research are considered promising routes in biotechnology for the biocontrol of foodborne pathogens and in human and animal therapy to fight bacterial resistance, but still have a long way to go before they reach the clinic. In the future, characterization techniques will be crucial to monitor phage production and modifications. Many other objects of prime biomedical importance are amenable to analysis by our technique. It is the case of other viruses like HIV or Zika, as well as cancer biomarkers like exosomes.

**References.**

**Acknowledgments.**

The authors acknowledge support from the LETI Carnot Institute NEMS-MS project, from the DGA Astrid NEMS-MS project, and from the European Union through the ERC Enlightened project (616251) and the Marie-Curie Eurotalents incoming (M.S., S.D.M.) fellowships. They would like to thank Dr. Virginie Brun for her support and fruitful discussions, as well as Dr. Anastasia Holovchenko. They thank Magali Aumont-Nicaise for her contribution in differential scanning calorimetry experiments. This work has benefited from the facilities and expertise of the Macromolecular Interaction Platform of I2BC (UMR 9198).

**Supplementary materials.**

Materials and Methods

Figs. S1 to S19